\documentstyle[12pt,epsfig]{article}
\newcommand{\be}{\begin{equation}}
\newcommand{\en}{\end{equation}}
\newcommand{\beeq}{\begin{eqnarray}}
\newcommand{\eneq}{\end{eqnarray}}

\makeatletter
\def\vereq#1#2{\lower3pt\vbox{\baselineskip1.5pt \lineskip1.5pt
\ialign{$\m@th#1\hfill##\hfil$\crcr#2\crcr\sim\crcr}}}
\makeatother
\begin{document}

\begin{titlepage}
\begin{flushright}
TUM-PH-619/06\\
KEK-TH-1069
\end{flushright}
\begin{center}
{\Large\bf
Split-SUSY versus SUSY GUTs
}
\end{center}
\vspace{1cm}
\begin{center}
Naoyuki {Haba}$^{(a),}$
\footnote{E-mail: haba@ph.tum.de \\
On leave of absence from
 Institute of Theoretical Physics, University of
 Tokushima, 770-8502, Japan}
and
Nobuchika {Okada}$^{(b,c),}$
\footnote{E-mail: okadan@post.kek.jp},
\end{center}
\vspace{0.2cm}
\begin{center}
${}^{(a)}${\it Physik-Department,
 Technishe Universit$\ddot{a}$t M$\ddot{u}$nchen, James-Franck Strasse,
 D-85748 Garching, Germany}
\\[0.2cm]
${}^{(b)}$ {\it Theory Division, KEK, Tsukuba 305-0801,
Japan}
\\[0.2cm]
${}^{(c)}$ {\it Department of Particle and Nuclear Physics,
The Graduate University \\
for Advanced Studies (Sokendai), Oho 1-1, Tsukuba,
Ibaraki 305-0801, Japan.}

\end{center}
\vspace{1cm}
\begin{abstract}


The gauge coupling unification is one of main motivations 
 in the split-SUSY scenario, and the existence of  
 the grand unified theories (GUTs) is assumed. 
We examine how to realize split-SUSY mass spectrum 
 in the context of GUTs 
 and find that the construction of split-SUSY GUTs 
 is by no means straightforward. 
With $R$-symmetry breaking sources in the GUT sector, 
 GUT particles play a role of the messengers 
 in the gauge mediation scenario 
 and their contributions to gaugino masses 
 can be sizable. 
We find the upper bound on soft scalar masses of 
 ${\cal O}(10^{10})$ GeV 
 from consistency for constructing the split-SUSY GUT. 
Also, we discuss the attempt 
 to construct $R$-symmetric GUT models.

\end{abstract}
\end{titlepage}
\newpage
\section{Introduction}


It is well known that the gauge hierarchy problem 
 in the Higgs sector of the standard model (SM) can be solved 
 by introducing supersymmetry (SUSY)~\cite{review}.
The minimally extended SUSY SM (MSSM) has an elegant feature, 
 namely, the unification of three gauge couplings 
 at the scale $M \sim 2 \times 10^{16} 
 \mbox{GeV}$\cite{unification}, 
 so that the existence of SUSY and also 
 SUSY Grand Unified Theories (GUTs) 
 have been widely believed. 
Also, SUSY seems to be necessary for the construction 
 of consistent string theories which include quantum gravity. 
However, experiments have not observed any SUSY particles yet. 
In addition, the predicted proton-decay by SUSY GUTs 
 has not been observed yet~\cite{Eidelman:2004wy}. 
Then, we might consider the possibility of 
 heavy SUSY particles as one option. 

The split supersymmetry (split-SUSY) scenario
 proposed in Refs.\cite{Split-SUSY, Split-SUSY2} 
 is just the case, 
 in which scalar masses are super-heavy 
 while fermion masses are maintained to be around  
 the weak scale, protected by the $R$ (chiral) symmetry. 
This split-SUSY soft mass spectrum can be simply obtained 
 by taking decoupling limit of the scalar particles 
 except the SM-like Higgs boson in the MSSM\cite{Split-SUSY, Split-SUSY2}.  
Therefore, in this scenario, 
 the nature is fine-tuned intrinsically and 
 SUSY is nothing to do with the gauge hierarchy problem 
 (for the related studies, see for examples, 
 Refs.\cite{{spspectrum}}-\cite{{ho}}). 
Nevertheless, SUSY still plays two important roles. 
One is for the gauge coupling unification. 
It has been found that the gauge couplings are 
 successfully unified at the GUT scale 
 even with the split-SUSY soft mass spectrum\cite{Split-SUSY}. 
The other is to provide the dark matter candidate 
 (neutralino) as usual\cite{Split-SUSY, Split-SUSY2}. 
In other words, these two facts are main motivations 
 for the split-SUSY scenario. 

%
%

As discussed in Refs.\cite{Split-SUSY, Split-SUSY2}, 
 it is nontrivial to keep the split-SUSY mass spectrum 
 under quantum corrections. 
First of all, contributions by the anomaly mediation
 (AMSB)\cite{AMSB1, AMSB2} should be concerned. 
In the split-SUSY scenario, soft SUSY breaking masses 
 for sfermions are super-heavy 
 so that SUSY breaking scale is very high. 
This means that vacuum expectation value (VEV) 
 of superpotential is also very large 
 in order to obtain a vanishing cosmological constant 
 in normal supergravity. 
In this case, F-term of the compensating multiplet is 
 very large, and thus the AMSB contributions 
 make gauginos very heavy. 
To switch them off, 
 the structure of almost no-scale supergravity\cite{almost} 
 is necessary\cite{Split-SUSY, Split-SUSY2}. 
In general, to avoid gauginos obtaining large masses, 
 the $R$-symmetry, which works as the chiral symmetry for gauginos, 
 plays a crucial role. 
It has been shown that, 
 using (approximate) $R$-invariance in the MSSM 
 and the hidden sectors, 
 the split-SUSY mass spectrum 
 can be realized\cite{Split-SUSY, Split-SUSY2}.  

However, since the grand unification is one of main motivations 
 of the split-SUSY scenario, 
 we must examine how to realize 
 the split-SUSY mass spectrum in the context of GUTs. 
In this paper, we address this issue and find that, 
 concerning GUT models, 
 to realize the split-SUSY spectrum is not straightforward. 
In a simple GUT model, there are some $R$-symmetry breaking sources 
 associated with new GUT particles and their VEVs 
 to break the GUT gauge symmetry down to the SM one. 
%
%
%
%
Then, there emerge some large contributions to 
 gaugino masses due to the GUT particles. 
To avoid these contributions, we must construct the GUTs 
 in which $R$-symmetry breaking effects 
 from $A$-terms, hidden sector fields 
 and the VEV of the superpotential are all small. 
These considerations lead to the upper bound 
 on soft scalar masses as ${\cal O}(10^{10})$ GeV. 
We also discuss attempt to construct $R$-symmetric GUTs (RGUTs). 
It seems to be very difficult to obtain a realistic RGUTs 
 maintaining the gauge coupling unification.

\section{Radiative corrections from GUT particles}

The split-SUSY is essentially 
 the MSSM with the decoupling of
 scalar fields except for one SM-like Higgs.
For the split-SUSY spectrum,
 we need the huge $F$-term
 for the heavy scalar masses. 
We introduce the gauge singlet field, $X$, 
 taking a $F$-term as
\begin{equation}
\label{X}
 X = F_X \theta^2.
\end{equation}
The split-SUSY requires the high scale SUSY breaking 
 as\cite{Split-SUSY, Split-SUSY2}.  
\begin{equation}
\label{spsp}
 {F_X \over M_p}\equiv \tilde{m} \simeq 10^6 \sim 10^{12}\: {\rm GeV},
\end{equation}
where $M_p$ denotes the reduced Planck scale. 
Scalar soft SUSY breaking masses are obtained as
\begin{equation}
 \int d^4 \theta {X^\dagger X \over M_p^2} Q^\dagger Q = \tilde{m}^2
                |\tilde{Q}|^2,
\end{equation}
where $Q$ denotes visible sector superfields. 
The large soft squared masses of sfermions and 
 Higgs are induced from this operator. 
Also, the so-called $B$-term is induced similarly as 
\begin{equation}
\label{Bterm}
 \int d^4 \theta {X^\dagger X \over M_p^2} (H_u {H_d} + h.c.)
    = \tilde{m}^2 (H_u {H_d} + h.c.),
\end{equation}
 where $H_u$ and $H_d$ denote the Higgs doublets. 
It suggests $B \mu \sim \tilde{m}^2$,
 which is needed for the structure of 
 the split-SUSY\cite{ho}.

Since the $F$-term in Eq.(\ref{X}) is huge, 
 the vanishing cosmological constant requires 
 the large constant superpotential, 
 so that gravitino mass is very large. 
To avoid large contributions to gaugino masses 
 due to the AMSB, 
 the structure of the almost no-scale supergravity\cite{almost} 
 should be incorporated, 
 which leads to a hierarchy between gravitino mass 
 and soft scalar masses, $m_{3/2} \gg \tilde{m}$ 
 (for basic formulas, see, for example, Ref.\cite{ho}). 
Furthermore, note that the superpotential 
 generally induces negative soft scalar mass squareds 
 of ${\cal O}(0.01 \times m_{3/2}^2)$ 
 through the gravitino one-loop diagram\cite{gr}. 
Thus, as a realistic scenario, 
 we consider the sequestering scenario in extra dimensions, 
 where the huge constant superpotential is located 
 at the different brane spatially separated 
 from the brane on which the GUT sector and 
 the hidden sector field $X$ reside. 
In this setup, the negative contributions 
 through the gravitino loop diagram are 
 enough suppressed by a factor $(L M_p)^{-1}$\cite{gr} 
 with a large $L$ being the distance between two branes. 
On the other hand, there is no correction 
 for gaugino masses from this gravitino loop diagram, 
 because of $R$-symmetry\cite{gravitino-loop}. 

Since the grand unification is one of main motivations 
 of the split-SUSY scenario, 
 we should examine how the split-SUSY mass spectrum 
 can be realized in the context of GUTs. 
As an example, let us consider the minimal $SU(5)$ GUT. 
The Higgs superpotential is given by
\begin{eqnarray}
\label{GUT}
 {\cal W} = \frac{1}{2} M tr \Sigma^2 + \frac{1}{3} tr \Sigma^3 
  + \bar{H}\Sigma H + 3 M H \bar{H} , 
\end{eqnarray}
 where $M \simeq 10^{16}$ GeV is the mass 
 around the GUT scale, 
 $\Sigma$ is the adjoint Higgs, 
 and $H$ and $\bar{H}$ are ${\bf 5}$ and ${\bf \bar{5}}$ 
 representation Higgs fields, respectively. 
Here, we have dropped dimensionless coupling constants, 
 for simplicity. 
The adjoint Higgs, $\Sigma$, breaks $SU(5)$ to 
 the SM gauge group through its VEV,  
\begin{equation}
 \langle \Sigma \rangle = diag.(2,2,2,-3,-3) M. 
\end{equation} 
Triplet-doublet (TD) splitting is assumed 
 to be achieved by the fine-tuning. 

If $R$-charge of $X$ is zero ($Q_R(X)=0$), 
 soft SUSY breaking $A$ and $B$-terms can be induced as 
\begin{eqnarray}
\label{A1}
{\cal L}_{soft} 
 = \int d^2 \theta {X \over M_p} {\cal W} 
=  \tilde{m} [M tr \Sigma^2 + tr \Sigma^3 +
   \bar{H}\Sigma H + 3 M H \bar{H}] ,
\end{eqnarray}
{}from Eq.(\ref{GUT}), 
 where all fields stand for scalar component. 
Immediately, we notice that 
 the gluino and bino obtain their masses 
 through 1-loop diagrams shown in Fig.1, 
 where colored-Higgs ($T, \bar{T}$) and colored-higgsino 
 ($\tilde{T}, \tilde{\bar{T}}$) are running in the loop. 
Here, the GUT particles play the role of the messenger 
 in the gauge mediated SUSY breaking\cite{GMSB}. 
In Fig.1, $M_T$ is the supersymmetric fermion mass 
 for $T\bar{T}$, 
 and $B_T$ is the SUSY breaking $B$-term 
 for $\tilde{T}\tilde{\bar{T}}$. 
These are given by $M_T \simeq M$ and $B_T \simeq \tilde{m}M$ 
 in the minimal $SU(5)$ GUT. 
Therefore, the 1-loop diagram makes masses 
 of gluino and bino huge, 
\begin{equation}
 M_{1/2}  \simeq {\alpha\over 4 \pi} {\tilde{m}M \over M} 
 \sim 10^{-2}\times \tilde{m}. 
\label{mM1}
\end{equation}
As for the wino mass, 
 there is a 1-loop diagram shown in Fig.2, 
 where the adjoint Higgs plays a role of the messenger. 
In the figure, the supersymmetric fermion mass of
 $\tilde{\Sigma}$ is denoted as $M_\Sigma$, 
 and scalar soft mass is denoted as $B_\Sigma$. 
Again, these are given by 
 $M_\Sigma \simeq M$ and $B_\Sigma \simeq \tilde{m}M$, 
 so that we obtain the same result as Eq.(\ref{mM1}). 
So, the wino also obtains too heavy mass.
Above contributions completely destroy 
 the hierarchical structure of the split-SUSY. 

\begin{figure}
\begin{center}
\epsfig{file=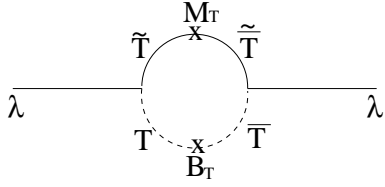,width=5cm}
\caption{Colored Higgs contribution to gaugino masses.}
\label{fig1}
\end{center}
\end{figure}
\begin{figure}
\begin{center}
\epsfig{file=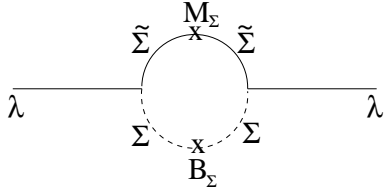,width=5cm}
\caption{Adjoint Higgs contribution to gaugino masses.}
\label{fig2}
\end{center}
\end{figure}

This difficulty seems to originate from the fact that 
 $R$-symmetry is broken badly by $F_X$ in Eq.(\ref{A1}). 
Reminding that we must impose $R$-symmetry 
 in order to construct the split-SUSY 
 spectrum, {\it fermion masses $\ll$ scalar masses},
 in the usual (non-GUT) MSSM, 
 we should avoid the large scale $R$-symmetry breaking. 
Thus, since the split-SUSY requires the large magnitude 
 of $F_X$ in Eq.(\ref{spsp}), we should take $Q_R(F_X)=0$. 
This implies $Q_R(X)=2$, so that 
 Eq.(\ref{A1}) is forbidden by the $R$-symmetry. 
Of course, since there is no $R$-symmetry in Eq.(\ref{GUT}) 
 from the beginning, we can, in general, add 
 any higher dimensional terms with $X$ 
 in spite of the $R$-charge assignment of $X$. 
In this paper, we impose $R$-symmetry 
 for SUSY breaking $A$ and $B$-terms, 
 the hidden sector field, and  
 higher dimensional terms among the GUT fields. 
For superpotential, we may omit some dangerous terms 
 for realization of split-SUSY mass spectrum 
 in the way of so-called ``technically natural'', 
 thanks to the non-renormalization theorem in SUSY theories.

In the split-SUSY scenario, 
 the weak-scale $\mu$-term is supposed to be realized 
 by the $R$-symmetry\cite{Split-SUSY, Split-SUSY2}, 
 the magnitude of $A$-term in Eq.(\ref{A1}) 
 should be suppressed and be replaced 
 to the weak scale\footnote{
For example, we introduce 
 an additional field, $Y$, 
 which is $Q_R(Y)=-2$. 
Then the $\mu$-term is given by 
$
 \int d^4 \theta \frac{X^\dagger Y}{M_p^2} (H_u {H_d} + h.c.)
    = \int d^2 \theta \: \delta \tilde{m} (H_u {H_d} + h.c.)
$.
Assuming $\langle Y \rangle /M_p \equiv \delta \: (\ll 1)$,
 $A$-terms in the GUT sector 
 as well as  $\mu$ term is 
 suppressed to the weak scale.
} 
as  
\begin{eqnarray}
{\cal L}_{soft} &=& \int d^2 \theta \: \delta \tilde{m}\: 
 \theta^2\: {\cal W} 
= \delta \tilde{m} [M tr \Sigma^2 + \lambda tr \Sigma^3 +
   \bar{H}\Sigma H + 3 M H \bar{H}] ,
\label{11}
\end{eqnarray}
where $\delta \tilde{m} \equiv \mu ={\cal O}(100)$ GeV. 
In this case, 
 scalar masses in Figs.1 and 2 are suppressed as
 $B_T \simeq B_\Sigma \simeq \mu M$, 
 so that the 1-loop correction to gaugino masses 
 from the GUT particles is given by 
\begin{equation}
 M_{1/2} \simeq {\alpha\over 4 \pi} {\mu M \over M} 
 \sim
 10^{-2}\times \mu, 
\label{suppress}
\end{equation}
which is negligibly small. 

The problem seems to be solved by imposing 
 $R$-symmetry as $Q_R(X)=2$. 
However, there exist other contributions. 
The adjoint Higgs and colored Higgs
 can obtain the soft masses through 
 the $D$-term as
\begin{eqnarray}
\label{Sigma}
&& \int d^4 \theta {X^\dagger X \over M_p^2} tr \Sigma^2 %
 = \tilde{m}^2    tr \tilde{\Sigma} ^2, \label{xxss} \\
&& \int d^4 \theta {X^\dagger X \over M_p^2} T \bar{T} 
 = \tilde{m}^2   \tilde{T} \tilde{\bar{T}}. 
\label{18}
\end{eqnarray}
These soft masses mean 
 $B_T \simeq B_\Sigma \simeq  \tilde{m}^2$, 
 which induce gaugino mass corrections as 
\begin{equation}
 M_{1/2}  \simeq {\alpha\over 4 \pi} {\tilde{m}^2 \over M} 
 \sim
 10^{-2} \times \tilde{m} 
 \left( \frac{\tilde{m}}{M}
 \right)
\label{15}
\end{equation} 
This implies the existence of the upper bound
 on scalar mass as 
 $\tilde{m} \leq 10^{10}$ GeV\footnote{
There are not so large phenomenologically
 suitable parameter region 
 in $10^{10} {\rm GeV} \leq \tilde{m} \leq 10^{12} 
 {\rm GeV}$\cite{Arvanitaki:2005fa}. 
}.  
This problem is originated from the fact that, 
 in the minimal $SU(5)$ GUT, 
 the $R$-symmetry is explicitly broken 
 in the superpotential Eq.(\ref{GUT}) from the beginning. 
Since non-vanishing gaugino masses need 
 both SUSY breaking and $R$-symmetry
 breaking, 
 the GUT scale $R$-symmetry breaking
 induces the serious problem.
This is the essential reason 
 for appearing the upper bound of $\tilde{m}$,
 when we extend the split-SUSY to the minimal $SU(5)$ GUT.

Here, we should comment on the effect of 
 non-zero F-component of $\Sigma$ ($F_\Sigma$). 
Once non-zero $F_\Sigma$ is developed, 
 $B_T$ and $B_\Sigma$ are produced through Eq.(\ref{GUT}).
Hence, gauginos obtain their masses as 
\begin{equation}
 M_{1/2}  \simeq {\alpha\over 4 \pi} {F_\Sigma \over M} . 
 \label{fsigma}
\end{equation} 
There are two possibilities for non-zero $F_\Sigma$ 
 to be produced after SUSY breaking as follows. 
One is a slight VEV shift by 
 $\langle \Sigma \rangle \simeq M 
 \rightarrow M(1+{\cal O}(\tilde{m}^2/M^2))$ 
 in the Higgs potential 
 after the soft mass term, $\tilde{m}^2 \Sigma ^2$, in 
 Eq.(\ref{Sigma}) is taken into account\footnote{
 The contribution from the soft mass,
 $\mu M tr \Sigma^2$ in Eq.(\ref{11}),
 is negligible, since the VEV shift from
 this soft mass is  $\langle \Sigma \rangle \simeq 
 M(1+{\cal O}(\mu M/M^2))$, which induces
 $F_\Sigma ={\cal O}(\mu M)$, and 
 the gaugino mass correction is given by 
 $M_{1/2} \simeq \frac{\alpha}{4\pi}\mu$.
}.
Then, we find $F_\Sigma ={\cal O}(\tilde{m}^2)$, 
 and the contribution from Eq.(\ref{fsigma}) 
 is of the same order of magnitude as 
 the one from Eq.(\ref{15}). 
%
%
%
%
The other possibility is 
 a tad-pole term for $F_\Sigma$ 
 in Eq.(\ref{xxss}) with non-zero VEVs 
 for $\langle X \rangle$, $F_X$ and $\langle \Sigma \rangle$. 
%
%
Together with $|F_\Sigma|^2$ from kinetic term, 
 non-zero $F_\Sigma$ is induced through equation of 
 motion for $F_\Sigma^\dagger$ such as 
\begin{equation}
\label{24}
 F_\Sigma \sim {\langle X^\dagger \rangle F_X \over M_p^2} 
 \langle \Sigma \rangle 
 \simeq \delta_X \tilde{m} M , 
\end{equation} 
where $\delta_X \equiv \langle X \rangle / M_p$. 
This parameter parameterizes 
 the magnitude of $R$-symmetry breaking 
 by the hidden sector field,  
 since $X$ has $R$-charge $Q_R(X)=2$. 
The value of $\delta_X$ must be small enough 
 to reproduce weak-scale gaugino masses.  
We should notice that 
 the value of $\langle X \rangle$ 
 depends on the hidden sector potential 
 and can be non-zero in general, 
 although we assume $\langle X \rangle = 0$ 
 in Eq.(\ref{X}). 

Here, let us briefly summarize discussions above. 
For the weak-scale gaugino masses, 
 the weak-scale $A$ and $B$-terms 
 are required 
 in the MSSM sector discussed in Refs.\cite{Split-SUSY, Split-SUSY2}. 
In addition to
 this approximate $R$-symmetry, 
 the following two conditions should be imposed  
 in the context of the minimal $SU(5)$ GUT with
 the split-SUSY structure, 
(i): $\tilde{m} \leq 10^{10}$ GeV,  
(ii): $\langle X \rangle \ll M_p$. 
%
%
%
\begin{equation}
\begin{array}{|c||c|c|}
\hline
\tilde{m}\:({\rm GeV}) 
 & \Delta M_{1/2}\:({\rm GeV}) & \delta_X \\
 & (\frac{\alpha}{4\pi}\frac{\tilde{m}^2}{M}) 
 & (\frac{\alpha}{4\pi}\delta_X \tilde{m} \leq 100\:{\rm GeV}) \\ \hline\hline
10^n &10^{2n-18} & \leq 10^{-(n-4)}\\ \hline
\end{array}
\end{equation}
Here, $n=6,7,\cdots, 12$.
The second condition is strongly dependent of the
 structures of the SUSY breaking hidden sector. 
We have shown that the split-SUSY GUT requires 
 $\tilde{m} \leq 10^{10}$ GeV 
 and $\delta_X \ll 1$ which means 
 the small $R$-symmetry breaking
 in the hidden sector.

The magnitude of the $R$-symmetry breaking
 in the hidden sector 
 parameterized by the magnitude of $\langle X \rangle$ 
 ($\delta_X$ in Eq.(\ref{24})), 
 is completely model dependent. 
When we consider the Polonyi hidden sector, ${\cal W}=m^2 X$, 
 as an example 
 in the framework of the almost no-scale supergravity, 
 we can easily show that 
 $\langle X \rangle =0$ is obtained 
 by introducing higher order terms 
 in the Kahler potential\cite{Split-SUSY2}, 
 ${\cal K}=X^\dagger X - (X^\dagger X)^2 + \cdots$. 
In order to obtain $\delta_X \ll 1$, 
 it is crucial that 
 the almost no-scale structure and the sequestering 
 of the constant superpotential mentioned above. 
In this setup, 
 potentials in the Polonyi sector and 
 the almost no-scale sector with the constant superpotential 
 on the other brane 
 works almost independently, 
 so that the vanishing cosmological constant can be achieved 
 between the positive contribution from the hidden sector 
 and the negative contribution 
 from the almost no-scale sector. 
Note that 
 if we consider the Polonyi hidden sector 
 in usual supergravity, 
 the vanishing cosmological constant condition 
 requires $X={\cal O}(M_p)$. 


Before closing this section, 
 we give further comments on our setup 
 in the framework of supergravity. 
After the GUT symmetry breaking, 
 Eq.(\ref{GUT}) leads to 
 $\langle {\cal W} \rangle ={\cal O}(M^3)$, 
 so that, in normal supergravity, 
 the huge SUSY breaking scale,  
 $F \simeq M^3/M_p \sim (10^{15} \:{\rm GeV})^2$, 
 is required to achieve 
 the vanishing cosmological constant. 
Therefore, soft scalar masses become 
 of order $\tilde{m}= F_X/M_p \sim 10^{12}$ GeV\footnote{
The gravitino mass becomes 
 also the same order
 as $m_{3/2}= \langle {\cal W} \rangle /M_p^2 
 \simeq F_X/M_p$.}, 
 which contradicts against the condition obtained above, 
 $\tilde{m} \leq 10^{10}$ GeV. 
This difficulty is also originated 
 from the GUT scale $R$-symmetry breaking 
 by $\langle {\cal W} \rangle$\footnote{
Generally, 
 the condition of the vanishing cosmological constant 
 in the GUTs 
 tends to need too large SUSY breaking\cite{Izawa:1997he}. 
}. 
On the other hand, in the framework 
 of the almost no-scale supergravity, 
 the contribution of $\langle {\cal W} \rangle$ 
 to vacuum energy is suppressed 
 by the special structure of the no-scale supergravity 
 and we can avoid this problem. 
However, a large $\langle {\cal W} \rangle$ induces 
 a large {\it negative} contribution 
 to scalar squared masses 
 through the gravitino loop diagram, 
 $ \tilde{m}^2 \sim - 0.01 |{\cal W}|^2/M_p^4$. 
For $\langle {\cal W} \rangle ={\cal O}(M^3)$, 
 this contribution is too large 
 $|\tilde{m}| ={\cal O}(10^{11})$ GeV 
 to be consistent with the condition we obtained. 
To avoid this problem, 
 we need to introduce a constant superpotential term, $-M^3$, 
 in the visible sector superpotential in Eq.(\ref{GUT}), 
 so as to achieve small VEV of the total superpotential 
 through the fine-tuning.  

\section{$R$-symmetric GUTs}



All the difficulties discussed in the previous section
 originates from the fact that
 the minimal $SU(5)$ GUT breaks $R$-symmetry
 explicitly at the GUT scale.
It implies that 
 the $R$-symmetry should not
 be broken at the GUT scale              
 for maintaining weak-scale gaugino masses.     
In this section, we discuss some attempt 
 to construct RGUTs\footnote{
The continuous global $U(1)_R$ symmetry 
 might not be necessary. 
Even a discrete symmetry $Z_{3R}$ 
 or larger might be useful for preserving 
 weak-scale gaugino masses. }. 
We consider two explicit models of RGUT as examples. 

The first model is the 5D $SU(5)$ GUT 
 in which there is no $\Sigma$ and 
 the GUT gauge group is broken through 
 the boundary condition\cite{Kawamura:2000ev,{Hall:2001pg},{Hebecker:2001jb}}.
Thus, there is no need to take into account 
 the diagrams of $\Sigma$ in Fig.2.
This model possesses the 
 $R$-symmetry, which is broken only by
 the weak-scale mass parameters, 
 $\mu$-term and gaugino masses. 
The gauge coupling unification, which 
 is another motivation in the split-SUSY, 
 can be achieved in this setup\cite{{Hall:2001pg}}.
We can also neglect
 the VEV of $\langle {\cal W} \rangle$,
 because of the absence of the adjoint field. 

As for the colored-Higgs contribution in Fig.1, 
 it is sufficiently suppressed as follows.
The $H$ and $\bar{H}$ 
 Higgs fields are in the bulk
 with 
 $R$-charge of $Q_R(T\bar{T})=0$. 
This is the reason why 
 the value of $\mu$ in the 
 $\mu T\bar{T}$ term in  ${\cal W}$ 
 is suppressed to the weak scale. 
This $R$-charge assignment
 allows 
 the soft scalar mass, $\tilde{m}^2 \tilde{T} \tilde{\bar{T}}$, 
 in Eq.(\ref{18}).
On the other hand,
 the effective fermion mass 
 becomes $M_T \simeq M^2/\mu$ in the diagram 
 in Fig.1, since 
 $T$ and $\bar{T}$ have the compactification 
 (GUT scale) masses with their chiral partner of 5D.
It works just like the missing partner models, 
 and the contribution through the diagram of Fig.1 
 is suppressed by an additional factor, $\mu/M$,
\begin{equation}
 M_{1/2}  \simeq {\alpha\over 4 \pi} {\mu \over M}
 {\tilde{m}^2 \over M}.
\end{equation} 
This is negligibly small. 

The split-SUSY in this 5D $SU(5)$ GUT
 was considered in Ref.\cite{{Sarker}}.
(The gauge coupling unification is
 discussed in Ref.\cite{{Sarker2}}.) 
%
%
However, since the gauge multiplets reside 
 in the bulk and couple to the radion multiplet,
 we have to consider a contribution to gaugino masses 
 due to the F-term of the radion ($F_T$), 
 (which is equivalent\cite{{Kaplan:2001cg}} to the mechanism 
 of the Scherk-Schwarz SUSY breaking\cite{Scherk:1978ta}). 
%
Through non-zero $F_T$, 
 gauginos obtain masses of order of gravitino mass 
 in the framework of the almost no-scale supergravity, 
 so that we cannot realize the split-SUSY mass spectrum. 
Thus, some modification of the setup is necessary. 
The simplest extension might be to consider 
 a six dimensional model with 
 the compactification of extra two dimensions 
 on the orbifold $(S^1/Z_2)^2$, and
 set the 5D GUT sector only on the 5th dimensional direction 
 and the almost no-scale sector in Ref.\cite{ho}
 on the 6th dimensional direction.  
However, in general, we will not able to 
 assume the sequestering between 
 almost no-scale sector and the 5D GUT sector, 
 because two radion fields corresponding 
 each extra dimensional radius 
 must couple with each other in a six dimensional supergravity. 
Thus, we may conclude that it is difficult 
 to construct the consistent split-SUSY GUT in extra dimensions. 

The second model of the RGUT is the 4D non-minimal 
 $SU(5)$ GUT\cite{{Izawa:1997he}},  
 in which 
 an additional
 adjoint superfield, $\Sigma'$,
 is
 introduced. 
The superpotential is extended as
\begin{equation}
\label{RGUT}
 {\cal W} = M tr( \Sigma \Sigma') + 
 tr (\Sigma^2 \Sigma' ) +
  \bar{H}\Sigma H + M H \bar{H} .
\end{equation}
The $R$-charges are
 $Q_R(\Sigma)=0$,
 $Q_R(\Sigma')=2$, and
 $Q_R(T\bar{T})=2$.
The $\Sigma'$ has vanishing VEV as    
 $\langle \Sigma' \rangle =0$
 in the SUSY vacuum. 
The VEV of the superpotential
 also vanishes $\langle {\cal W}\rangle =0$,
 so that $R$-symmetry preserved
 at the GUT scale. 
The TD splitting is assumed to be realized by the 
 fine-tuning\footnote{
 The main discussion in
 Ref.\cite{Izawa:1997he} is
 not the Eq.(\ref{RGUT}). 
They considered  
 the extra $SU(3)_H$ to achieve
 the TD splitting. 
We do not take this model, since 
 we would like to consider 
 the simple unification
 of the three gauge couplings. }. 


In this model, 
 the soft scalar mass of $\Sigma^2$
 is $\tilde{m}^2$ as Eq.(\ref{Sigma}),
 while that of $\Sigma'^2$ is
 suppressed due to the $R$-symmetry, since 
 we can not write down 
 the operator of Eq.(\ref{Sigma})
 for $\Sigma'^2$. 
We have shown in the previous section that 
 the $\mu$-term should be suppressed 
 to the weak scale by the $R$-symmetry. 
In the same way, 
 the magnitude of the soft mass
 of $\Sigma'^2$ should be suppressed 
 at least ${\cal O}(\mu M)$.
The soft squared mass term, 
 $\tilde{m}^2\Sigma^2$,
 causes the slightly shift of VEV as
 $\langle \Sigma \rangle \simeq M
 \rightarrow M(1+{\cal O}(\tilde{m}^2/M^2))$.
There is no VEV shift for $\langle \Sigma' \rangle$. 
This is because the SUSY vacuum of $\Sigma'$ 
 exists at the origin, 
 so that the vacuum is not shifted from
 the origin by the soft SUSY breaking parameters. 
However, here
 we assume the possibly maximal shift 
 of $\langle \Sigma' \rangle = {\cal O}(\mu)$
 concerning to some other effects of SUSY breaking.  
These VEV-shifts produce
 $\langle F_\Sigma \rangle \simeq \mu M$
 and 
 $\langle F_\Sigma' \rangle \simeq \tilde{m}^2$.

At first,
 we estimate the contributions from
 Fig.2. 
The scalar soft mass is $B_\Sigma \simeq \tilde{m}^2$ as 
 shown above (Eq.(\ref{Sigma})). 
We can show that  
 $\langle F_\Sigma' \rangle \simeq \tilde{m}^2$
 also induces the scalar mass of $B_\Sigma \simeq \tilde{m}^2$. 
The supersymmetric fermion 
 mass, $M_\Sigma \simeq \mu$, should be 
 induced through the $R$-symmetry
 breaking just like the $\mu$-term. 
Then the contribution from the 1-loop diagram in 
 Fig.2 becomes
\begin{equation}
 M_{1/2}  \sim {\alpha\over 4 \pi} {\mu \tilde{m}^2 \over \tilde{m}^2}
 \sim {\alpha\over 4 \pi} {\mu},
\end{equation} 
which is negligibly small.

Next, we show the contribution from the
 diagram in Fig.3. 
The soft squared scalar mass of it should be
 given by $B_{\Sigma}' \simeq \mu \tilde{m}$ maximally,
 since $Q_R(\Sigma'^2)=4$ suggests 
 no contribution from Eq.(\ref{Sigma}). 
Without $R$-symmetry breaking,
 this term can not be induced,
 and it is suppressed at least 
 by ${\cal O}(\mu \tilde{m})$.
The fermion mass of $M_\Sigma' \simeq \mu$  
 is induced through the $R$-symmetry
 breaking just like the $\mu$-term. 
Then, the 1-loop diagram from
 Fig.3 becomes
\begin{equation}
 M_{1/2}  \simeq {\alpha\over 4 \pi} {\mu^2 \tilde{m}  \over \mu \tilde{m}}
 \sim {\alpha\over 4 \pi} {\mu},
\end{equation} 
which is also negligible.

\begin{figure}
\begin{center}
\epsfig{file=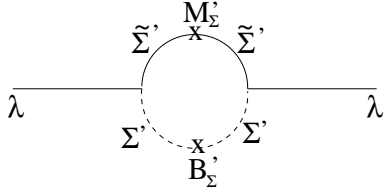,width=5cm}
\caption{$\Sigma'$ contribution to gaugino masses.}
\label{fig3}
\end{center}
\end{figure}

The third 
 is 
 contribution from the diagram in Fig.4.
The $R$-symmetry
 suppresses the SUSY breaking
 $B$-term as 
 $B_{\Sigma\Sigma'}\simeq \mu M$.
$\langle F_\Sigma \rangle \simeq \mu M$ from
 the VEV shift
 also induces 
 the same magnitude of
 $\mu M \Sigma\Sigma'$.
Since the supersymmetric
 fermion mass is $M$, 
 the 1-loop diagram from
 Fig.4 induces 
\begin{equation}
 M_{1/2}  \simeq {\alpha\over 4 \pi} {\mu M \over M}
 \simeq {\alpha\over 4 \pi} {\mu},
\label{22}
\end{equation} 
which is also tiny. 

\begin{figure}
\begin{center}
\epsfig{file=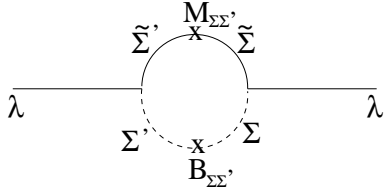,width=5cm}
\caption{$\Sigma, \Sigma'$ contribution to gaugino masses.}
\label{fig4}
\end{center}
\end{figure}

Finally we estimate the contribution from
 $T,\bar{T}$ in Fig.1. 
The soft scalar mass Eq.(\ref{18}) is suppressed
 by the $R$-symmetry $(Q_R(T\bar{T})=2)$,
 which should be estimated 
 to be maximally 
 ${\cal O}(\mu M)$.
The $A$-term suppressed by the $R$-symmetry 
 (Eq.(\ref{11})) and 
 the VEV shift induced $\langle F_\Sigma \rangle ={\cal O}(\mu M)$
 also suggest 
 $B_T \simeq \mu M$. 
Since the supersymmetric 
 fermion mass is $M_T \simeq M$,
 the 1-loop diagram from
 Fig.1 induces the
 same negligible correction as Eq.(\ref{22}).   
Therefore, the 1-loop diagrams contributions from 
 the GUT particles 
 are negligibly small in this model. 

The $R$-symmetry breaking from 
 $\langle {\cal W} \rangle$ is estimated as
\begin{equation}
 \langle {\cal W} \rangle \sim \mu M^2 
 \sim 10^2 \times (10^{16})^2 \; {\rm GeV}^3 . 
\end{equation} 
This is small enough 
 for contributions through the gravitino loop diagram 
 to be negligible. 

%
%
%

Although the above model works very well 
 to obtain split-SUSY mass spectrum, unfortunately,
 colored doublet components of the $\Sigma'$ superfields
 are light, so that the gauge coupling unification,
 which is one of the strongest motivations 
 of the split-SUSY, is destroyed. 
They can get masses through 
 the $R$-symmetry breaking term, 
 $\mu tr \Sigma'^2$, in Eq.(\ref{RGUT}).
This term should naturally be the weak scale.
In this case, 
 the VEV shift of $\langle \Sigma' \rangle ={\cal O}(\mu)$
 might be happened in the direction 
 which breaks color. 
Therefore, this mass term must be larger 
 than the weak scale. 
Even if we assume it, 
 the gauge coupling unification
 does not achieved, 
 unless this mass term is around the GUT scale, 
 which is quit unnatural. 
In general, there remains the flat direction 
 in RGUT\cite{Barr:2005xy}, 
 which implies the existence of additional light fields. 
Thus, unfortunately, we should extend 
 the matter content of the MSSM to reproduce 
 the gauge coupling unification in 4D RGUT. 
It is nontrivial whether such model can be 
 constructed within the framework of the RGUT models. 
Even if such a model can be successfully constructed, 
 the original motivation of the split-SUSY scenario 
 will be broken. 



\section{Summary}

Recently, the split-SUSY scenario has been proposed, 
 in which scalar soft masses are all super-heavy 
 while fermion soft masses are around the weak scale. 
This mass spectrum is obtained 
 by taking decoupling limit 
 of the scalar particles (except the SM like Higgs) 
 in the MSSM.  
It has been shown that, even with this split-SUSY mass spectrum, 
 the gauge couplings are successfully unified, 
 which implies the existence of the GUT behind the MSSM. 
Thus, the grand unification is one of main motivations 
 of the split-SUSY scenario. 

Although realization of the split-SUSY mass spectrum 
 have been discussed in the context of the MSSM, 
 it is necessary to examine it in the context of GUTs 
 because the existence of the GUT is one of 
 the most important assumptions in the split-SUSY scenario. 
We have shown that, if we consider the GUT sector, 
 GUT fields play a role of the messengers 
 in the gauge mediated SUSY breaking 
 and induce sizable contributions to gaugino masses, 
 so that the construction of the split-SUSY GUT 
 is not straightforward. 
We have considered 
 many possible contributions 
 which destroy the split-SUSY mass spectrum 
 and obtained the upper bound 
 on soft scalar masses as ${\cal O}(10^{10})$ GeV. 
Furthermore, 
 noting that dangerous contributions 
 originated from the $R$-symmetry breaking 
 in the GUT sector, 
 we have discussed the attempt to construct 
 $R$-symmetric GUT models in both four dimensions 
 and extra dimensions. 
Although the $R$-symmetric $SU(5)$ GUT model works well 
 to preserve the split-SUSY mass spectrum, 
 additional light fields 
 (which seems to be associated with $R$-symmetry) 
 appear in the model and 
 the success of the gauge coupling unifications 
 is destroyed. 
Thus, we need to extended the matter sector to reproduce 
 the gauge coupling unification. 
Such an extention of the MSSM sector 
 destroys the original motivation of the split-SUSY scenario. 
The contributions from the 
 GUT particles disturb the realization of the 
 approximate $R$-symmetry in the low energy, 
 and the construction of 
 the split-SUSY spectrum in the context of GUTs 
 is much more nontrivial than that in the MSSM.

\vskip 1cm

\leftline{\bf Acknowledgments}
We would like to thank A. Romanino 
 for fruitful and useful discussions.
NH is supported by Alexander von Humboldt Foundation.
This work is supported in part by Scientific Grants from
 the Ministry of Education and Science in Japan
 (Grant Nos.\ 15740164, 16028214, and 16540258).



\vspace{5mm}

\end{document}